\documentclass[twocolumn,amsfonts,showpacs,superscriptaddress,nofootinbib]{revtex4-1}
\usepackage{graphicx}
\usepackage{amssymb}
\usepackage{bm}
\usepackage{color}

\newcommand{\bc}{\begin{center}}
\newcommand{\ec}{\end{center}}
\def\ba#1{\begin{array}{#1}\displaystyle}
\newcommand{\ea}{\end{array}}

\newcommand{\beq}{\begin{equation}}
\newcommand{\eeq}{\end{equation}}
\newcommand{\beqa}{\begin{eqnarray}}
\newcommand{\eeqa}{\end{eqnarray}}
\newcommand{\no}{\nonumber}
\newcommand{\n}{\nonumber\\}
\newcommand{\bi}{\begin{itemize}}
\newcommand{\ei}{\end{itemize}}

\def\lt#1{\left#1}
\def\rt#1{\right#1}
\def\t#1{\tilde{#1}}
\def\h#1{\hat{#1}}

\def\frc#1#2{\frac{#1}{#2}}

\newcommand{\p}{\partial}

\newcommand{\Z}{{\mathbb{Z}}}

\newcommand{\R}{{\mathbb{R}}}

\newcommand{\Or}{{\cal O}}

\newcommand{\ep}{\epsilon}
\newcommand{\varep}{\varepsilon}

\newcommand{\Tr}{{\rm Tr}}

\newcommand{\dd}{{\rm d}}

\newcommand{\vtr}{v_{\rm tr}}
\newcommand{\qq}{{\tt q}}
\newcommand{\jj}{{\tt j}}
\newcommand{\uu}{{\tt k}}
\newcommand{\hh}{{\tt h}}
\newcommand{\pp}{{\tt p}}
\newcommand{\ww}{{\tt b}}
\newcommand{\cc}{{\tt c}}
\newcommand{\TT}{{\tt T}}
\newcommand{\rl}{{\rm l}}
\newcommand{\rr}{{\rm r}}
\newcommand{\rs}{{\rm stat}}
\def\sta#1{\langle #1\rangle_{\rm stat}}
\def\ini#1{\langle #1\rangle_{0}}
\def\gibl#1{\langle #1\rangle_{\rl}}
\def\gibr#1{\langle #1\rangle_{\rr}}

\def\gen#1{\langle #1\rangle}
\def\genc#1{\langle #1\rangle^{{\rm c}}}
\def\ve#1{\underline{#1}}

\def\eqref#1{(\ref{#1})}

\begin{document}

\title{Lower bounds for ballistic current and noise in non-equilibrium quantum steady states}

\author{Benjamin Doyon}
\affiliation
{Department of Mathematics, King's College London, Strand WC2R 2LS, UK}


\begin{abstract}
Let an infinite, homogeneous, many-body quantum system be unitarily evolved for a long time from a state where two halves are independently thermalized. One says that a non-equilibrium steady state emerges if there are nonzero steady currents in the central region. In particular, their presence is a  signature of ballistic transport. We analyze the consequences of the current observable being a conserved density; near equilibrium this is known to give rise to linear wave propagation and a nonzero Drude peak. Using the Lieb-Robinson bound, we derive, under a certain regularity condition, a lower bound for the non-equilibrium steady-state current determined by equilibrium averages. This shows and quantifies the presence of ballistic transport far from equilibrium. The inequality suggests the definition of ``nonlinear sound velocities'', which specialize to the sound velocity near equilibrium in non-integrable models, and ``generalized sound velocities'', which encode generalized Gibbs thermalization in integrable models. These are bounded by the Lieb-Robinson velocity. The inequality also gives rise to a bound on the energy current noise in the case of pure energy transport. We show that the inequality is satisfied in many models where exact results are available, and that it is saturated at one-dimensional criticality.
\end{abstract}

\maketitle

\section {Introduction and main results} Quantum systems out of equilibrium present some of the most important challenges of current theoretical physics. A family of non-equilibrium states which offer the hope of a deeper understanding and a stronger framework are quantum steady states carrying flows of energy, charge or particles with constant rates. These are the simplest non-equilibrium states, yet display many aspects of physics far from equilibrium, including non-equilibrium fluctuation relations and entropic fluctuations \cite{gallavotti,jarzynski,esposito,pillet}. They are of particular interest within many-body systems, where the interplay between quantum behaviours and non-equilibrium physics is prominent. In order to develop a general theory of non-equilibrium quantum steady states, it is of paramount importance to obtain further model-independent results. A fundamental problem is to establish conditions for the existence of non-equilibrium currents, and quantify these currents and their cumulants. The aim of the present paper is to make progress in this direction. We derive a lower bound, expressed in terms of equilibrium averages, for non-equilibrium ballistic currents and noise in local many-body systems.

\begin{figure}
\includegraphics[width=7cm,height=3.8cm]{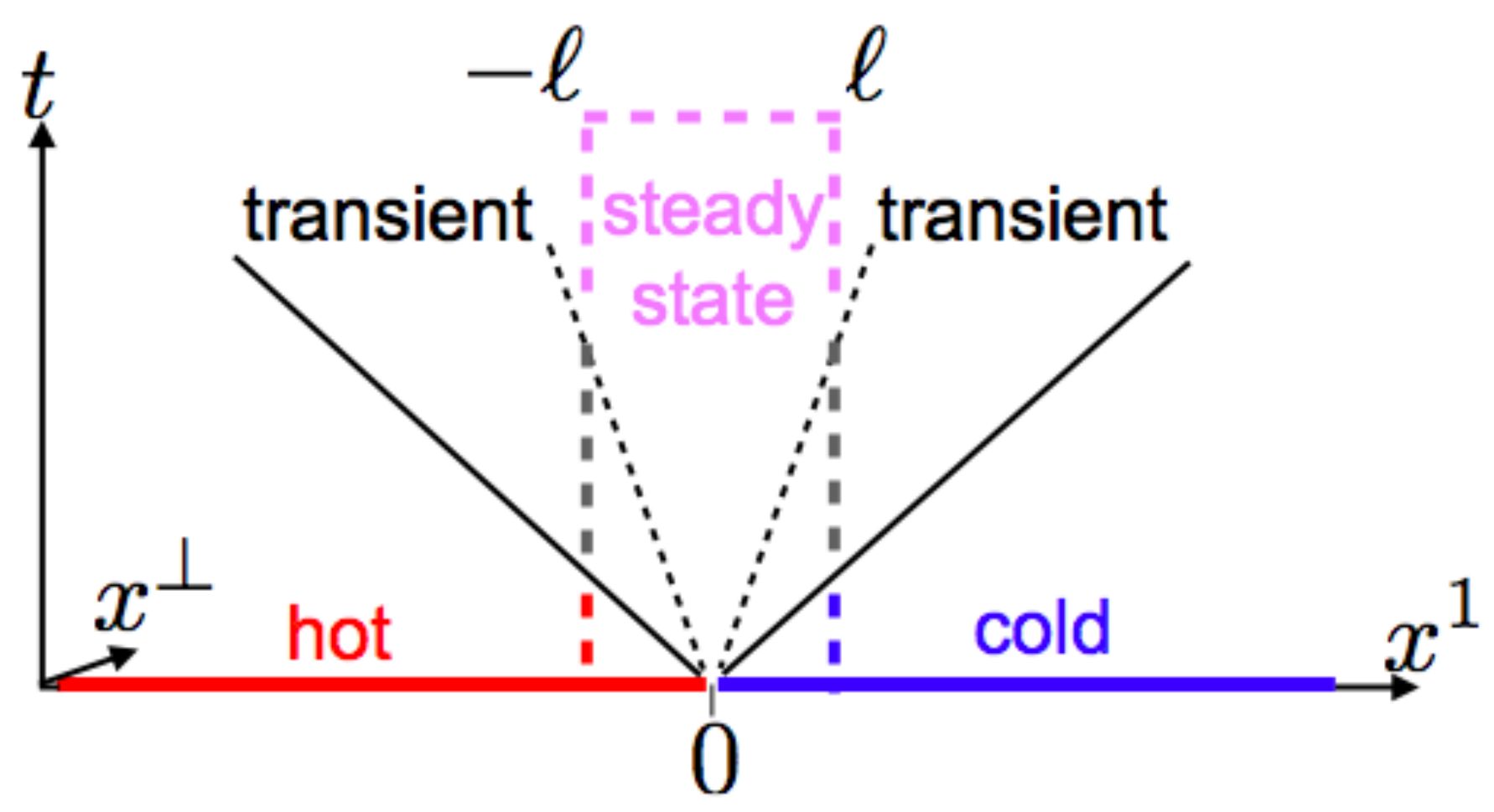}
\caption{The partitioning approach, in the example of a temperature difference. Transient regions separate asymptotic reservoirs from the central, steady-state region. The thick dotted line is the integration contour used in the proof.}
\label{quench}
\end{figure}

A construction of non-equilibrium steady states that takes into account the full unitary dynamics is the {\em partitioning approach}, introduced in \cite{caro,rub,spo} in one-dimensional models. Here we consider the general higher-dimensional setup studied in various recent works \cite{doyontrans,bhaseen1,cky,collura2,doyonKG}. Two halves, the left and right (with longitudinal coordinate $x^1\gtrless0$), of an infinite $d$-dimensional homogeneous system are independently thermalized, then connected and unitarily evolved for a long time. In this construction, the asymptotic regions  $x^1\to\pm\infty$ play the roles of reservoirs, furnishing and absorbing energy, particles or charge (see Fig. \ref{quench}). Generically, the steady state resulting at infinite times in the central region does not carry currents because of diffusive effects. However, non-equilibrium currents are expected to emerge if ballistic transport is allowed by the dynamics. This is of special interest for many-body physics, as ballistic transport points to anomalous, often collective, behaviours.

Assume that a longitudinal current observable $\jj:=\ve\jj^1$ is the density of a conservation law,
\beq\label{JI}
	\p_t\jj + \nabla\cdot \ve\uu =0.
\eeq
The current usually describes the transfer of a quantity $\qq$ (energy, charge or particle density) via another conservation law $\p_t \qq+\nabla\cdot\ve\jj=0$, and $\uu:=\ve\uu^1$ is interpreted as the ``longitudinal pressure''. Near equilibrium and under the assumption of local Gibbs thermalization, these conservation laws give rise to linear wave equations (``sound waves'') determined by the equations of state, whence to ballistic transport of small perturbations. Formally, a related statement is that relation \eqref{JI} leads to a nonzero Drude peak \cite{maz,zot}. One may ask to what extent such ballistic transport subsists far from equilibrium.

We will show that under a certain condition of regularity for the dynamics of $\uu$ in transient regions, there is also {\em non-equilibrium} ballistic transport if \eqref{JI} holds. The non-equilibrium stationary value of the current $\sta\jj$, in the partitioning approach, is bounded from below by the difference of {\em equilibrium} bulk averages of $\uu$ in the states of the original left and right halves, $\gibl\uu$ and $\gibr\uu$ respectively. Taking without loss of generality a current flowing from the left to the right,
\beq\label{jineq}
	v\sta\jj \geq \frc{\gibl\uu-\gibr\uu}2.
\eeq

The only model-dependent parameter in \eqref{jineq} is $v$, the Lieb-Robinson velocity \cite{lr,nach}. This is a fundamental characteristic of many-body quantum systems representing the maximum propagation velocity of information. The associated Lieb-Robinson bound, at the basis of our derivation, has received renewed interest recently, giving rise to impressive general results  \cite{hastlsm,hast,sims,brav,eis1,eis2,nach}. In the relativistic field theory (scaling) limit, the Lieb-Robinson velocity specializes to the velocity of light of the relativistic dispersion relation. The content of inequality \eqref{jineq} is that the pressure pushes the transferred quantities within regions where the information of the connection hasn't yet reached, thus contributing to the current via equilibrium averages as per the right-hand side of \eqref{jineq}; and may continue contributing within the transient regions in a way that is determined by the precise dynamics, thus giving rise to the inequality. With relativistic invariance, we will also derive thermodynamic relations leading to explicitly positive expressions for $\gibl{\uu}-\gibr\uu$ (see \eqref{relener}).

Inequality \eqref{jineq} has interesting consequences and interpretations. First, we may define a {\em transient velocity}
\beq\label{veff}
	\vtr := \frc{\gibl\uu - \gibr\uu}{2\sta\jj},
\eeq
and inequality (\ref{jineq}) says that this is bounded by the Lieb-Robinson velocity, $\vtr\leq v$. We may also define left and right transient velocities $v_{\rl,\rr}$ whose arithmetic average give $\vtr$,
\beq\label{vlr}
	v_{\rl,\rr} := \pm\frc{\gen{\uu}_{\rl,\rr} - \sta\uu}{\sta\jj},\quad
	\vtr = \frc{v_\rl+v_\rr}2,
\eeq
and we will show the stronger statements  $v_{\rl}<v$ and $v_\rr<v$. Transient velocities characterize the transient regions, and have the interpretation as velocities of effective sharp world lines or hypersurfaces separating asymptotic reservoirs from the steady state region.

Near equilibrium (in the limit where the left and right subsystems are in the same initial state), within linear response, we will show that, in {\em non-integrable systems}, $\vtr$ and $v_{\rl,\rr}$ all specialize to the sound velocity $v_{\rm s}$ associated with the wave equation linked to \eqref{JI},
\beq\label{limv}
	\lim_{\text{equilibrium}} v_{\rl,\rr} = v_{\rm s}.
\eeq
Hence the transient velocities also have the interpretation as {\em non-linear sound velocities}, describing non-equilibrium wave propagation beyond the linear response regime. Interestingly, relation \eqref{limv} does not hold in integrable systems. This is due to the lack of Gibbs thermalization: in the presence of infinitely many conservation laws, local thermalization ideas \cite{deu,sre,rigolth} are extended to generalized Gibbs ensembles \cite{Kinoshita,rigol,rigol2}. This precludes the emergence of the usual wave equation, based on Gibbs thermalization, even near equilibrium. In this case, the limits on the left-hand side of \eqref{limv} can be interpreted as {\em generalized sound velocities}, encoding generalized Gibbs thermalization. In particular, as we will see, these may take higher values than the usual sound velocity.

Second, from \eqref{jineq} one can derive an inequality for differential conductivities. Let $\eta$ parametrize a path on the space of equilibrium states, asymptotic reservoirs being at $\eta=\eta_{\rl,\rr}$ with $\eta_\rl\geq\eta_\rr$. Then we may define the differential conductivities $G_{\rl,\rr} = \pm \p\sta\jj / \p\eta_{\rl,\rr}$. Near enough to equilibrium (i.e. for all $|\eta_\rl-\eta_\rr|$ small enough), we have
\beq\label{Gineq}
	vG_{\rl,\rr} \geq \frc12 \frc{\p\gen{\uu}_{\rl,\rr}}{\p\eta_{\rl,\rr}}.
\eeq
For instance, for pure energy transport one may take $\eta$ to be the temperature. In this case, we can further show that \eqref{Gineq} implies a bound for the {\em noise} $c_2:=\int \dd t\, \dd^{d-1} x^\perp\,\lt(\sta{\jj(x^\perp,t) \jj(0)} - \sta{\jj}^2\rt)$ where $x^\perp$ is the coordinate transverse to the flow. With $\beta_{\rl,\rr}$ the inverse temperatures, for all $|\beta_\rl-\beta_\rr|$ small enough we have
\beq\label{c2}
	vc_2 \geq -\frc{1}2 \lt(\frc{\p}{\p\beta_{\rl}}\, \gen{\uu}_{\beta_{\rl}}
	+ \frc{\p}{\p\beta_{\rr}}\, \gen{\uu}_{\beta_{\rr}}\rt).
\eeq
This follows from the {\em extended fluctuation relations} (EFR)  \cite{doyontrans}, which relate higher fluctuation moments to derivatives of the average current. As we will argue based on \cite{doyontrans,bhaseen1}, this should hold generally for ballistic transport.

For the inequality \eqref{jineq} to hold we require the conservation law \eqref{JI} and a certain regularity condition.

The conservation law \eqref{JI} arises in many situations. In some integrable models, such as lattices of harmonic oscillators or the Ising, XY or XXZ (anisotropic Heisenberg) quantum chains, the energy current is the density for a nontrivial conserved charge, hence \eqref{JI} holds. Further, near or at quantum critical points, universal emerging collective behaviours ``wash out'' the lattice structure. With unit dynamical exponent, relativistic invariance emerges and the low-energy regime is described by Poincar\'e invariant quantum field theory (QFT) or by conformal field theory (CFT). Energy transfer observables are elements of the symmetric stress-energy tensor $\TT^{\mu\nu}$, with energy density $\qq = \TT^{00}$, current $\jj=\TT^{01}=\TT^{10}$ and (contravariant) pressure $\uu = \TT^{11}$  (the physical pressure is $\TT^{11}/v^2$), so that in particular current conservation \eqref{JI} follows from $\p_\mu \TT^{\mu\nu}=0$. The corresponding energy waves are sometimes called ``cosmic sound'' (these are studied for instance in  \cite{land}). Graphene provides a striking example, in which ballistic energy transport and its experimental significance were recently emphasized \cite{phan}. Thermal waves called ``second sound'', occurring from collective behaviours of phonon instead of electrons, are also known to occur in certain media \cite{smith}. In one-dimensional CFT, charge currents also satisfy \eqref{JI} thanks to chiral factorization.

The regularity condition, stated in Section \ref{sectproof}, is natural and expected to be widely valid, although generically hard to verify. It is satisfied, for instance, whenever there is large-scale monotonicity of $\uu$ in the transient regions. Conversely, its breaking, the knowledge of which only requires that of the stationary current and of equilibrium values of $\uu$, implies irregularity of $\uu$ in transient regions, hence large-scale non-monotonicity, a dynamical information otherwise hard to access.

Exact currents in the partitioning approach have been obtained both at and away from integrability and criticality: exact energy and charge currents in one-dimensional CFT \cite{gut,gut2,bd1,bd2}; exact energy currents in quantum chains with free-fermion representations \cite{tas,ah,og,aschp,delucaising} and in higher-dimensional free-field models \cite{collura2,doyonKG}, numerical observations \cite{Moore,Moore2,DeLucaVitiXXZ} in the XXZ chain, and proposed expressions in integrable QFT \cite{doyonint}, in (non-integrable) higher-dimensional CFT \cite{bhaseen1,cky} and in the XXZ chain \cite{DeLucaVitiXXZ}.  Using these available results, we verified the bound (\ref{jineq}) in many cases. We observe that it is saturated at one-dimensional criticality, strict in other cases, but broken by certain expressions from integrability at high enough temperatures.

Ballistic currents in homogeneous quantum systems are of interest not only as paradigmatic examples of non-equilibrium steady states, but also through their involvement in the theoretical formulation of more general quantum steady states. One way of representing reservoirs that treats effects of memory and entanglement exactly is via the partitioning approach: large quantum systems are initially thermalized, then connected to a small system through which flows are studied, the whole being unitarily evolved. This is a formulation used in the context of quantum dots \cite{ram,meir,hers,jauho,doyand}. However, such reservoirs will furnish thermalized carriers only if they are ballistically transported within them. The bound derived here provides a sufficient condition for this requirement, which in particular is fulfilled in the simple models of reservoirs used for quantum dots.

The rest of the paper is organized as follows. In Section \ref{sectnoneq}, we describe the setup. In Section \ref{sectbal}, we explain the relation between the partitioning approach and ballistic transport, and provide a non-equilibrium definition for the latter. In Section \ref{sectproof}, we state the regularity condition and provide the proof of \eqref{jineq}. In Section \ref{sectsound}, we discuss the transient and show \eqref{limv}. In Section \ref{sectmod} we analyze the inequality \eqref{jineq} within various models. In Section \ref{sectrel} we analyze the consequences of relativistic invariance. In Section \ref{sectnoise} we derive the inequality \eqref{c2} for the noise. Finally, we conclude in Section \ref{sectcon}.

\section{ Non-equilibrium quantum steady states}\label{sectnoneq}

For simplicity we use a continuous notation reminiscent of field theory (the results hold in lattices as well), and we set the Lieb-Robinson velocity to $v=1$. Let $\hh(x)$ be the energy density, at position $x=(x^1,x^\perp)$, for a homogeneous quantum system of dimension $d$. We separate the system into two reservoirs with commuting Hamiltonians $H_\rl$ and $H_\rr$:
\beq
	H_\rl = \int_{-L}^{0} \dd x^1\,\t\hh(x^1),\quad
	H_\rr = \int_0^{L} \dd x^1\,\t\hh(x^1)
\eeq 
where the longitudinal energy density is $\t\hh(x^1)=\int \dd^{d-1} x^\perp \,\hh(x^1,x^\perp)$, the integral of the energy density over the infinite $(d-1)$-dimensional transverse direction (for $d=1$, we have $\t \hh = \hh$ and use $x^1=x$). There is an interaction between the reservoirs on the $(d-1)$-dimensional flat transverse hypersurface at longitudinal coordinate $x^1=0$ (for $d=1$, this is a point at $x=0$), with Hamiltonian $\delta H_{\rl\rr}$, in such a way that the full Hamiltonian
\beq
	H=H_\rl+\delta H_{\rl\rr}+H_\rr = \int_{-L}^L \dd^d x\,\hh(x)
\eeq
is homogeneous.

In the partitioning approach, the initial density matrix  is thermal in the independent reservoirs,
\beq
	\rho_0 = \t \rho_{\rl}\otimes \t \rho_{\rr},\quad
	\t \rho_{\rl,\rr} = e^{-\beta_{\rl,\rr} \lt(H_{\rl,\rr}-\sum_j\mu_{\rl,\rr}^{(j)}Q_{\rl,\rr}^{(j)}\rt)}
\eeq
where $\mu_{\rl,\rr}^{(j)}$ are chemical potentials and $Q_{\rl,\rr}^{(j)}$ are conserved quantities with respect to $H_{\rl,\rr}$. Then, $\rho_0$ is evolved unitarily with the full Hamiltonian, and energy, charge or particles start flowing from one reservoir to the other. A non-equilibrium steady state may be reached after a long time, in a system that is large enough so that its reservoirs may provide and absorb an unbounded amount of thermalized quantities. Accordingly, the steady state, if it exists, is the limit
\beq
	\sta{\cdot} = \lim_{t\to\infty}\lim_{L\to\infty} \ini{e^{iHt}\,\cdot\,e^{-iHt}}
\eeq
where $\ini{\cdot} = \Tr\lt(\rho_0 \;\cdot\rt)/\Tr\lt(\rho_0\rt)$.
One expects the steady state to exist for families of observables supported on finite regions around the hypersurface $x^1=0$, for instance any local observable.

Without loss of generality, we assume that $Q_{\rl,\rr}^{(j)}$ give rise to corresponding bulk conserved quantities. That is, for $\delta Q_{\rl\rr}^{(j)}$ supported at $x^1=0$,
\beq\label{Qj}
	Q^{(j)} := Q_\rl^{(j)} + \delta Q_{\rl\rr}^{(j)} + Q_\rr^{(j)}
	= \int_{-L}^L \dd^dx\,\qq^{(j)}(x)
\eeq
are homogeneous conserved quantities with respect to $H$. The full Hamiltonian $H$ and conserved charges $Q^{(j)}$ give rise to bulk averages $\gen{\cdot}_{\rl,\rr}$ as occurring in \eqref{jineq}:
\beq
	\gen{\cdot}_{\rl,\rr} = \frc{\Tr\lt(\rho_{\rl,\rr} \cdot\rt)}{\Tr\lt( \rho_{\rl,\rr}\rt)},\quad
	\rho_{\rl,\rr} = e^{-\beta_{\rl,\rr} \lt(H-\sum_j\mu_{\rl,\rr}^{(j)}Q^{(j)}\rt)}.
\eeq

The above could be generalized to include an additional quantum system between the reservoirs: in this case $\delta H_{\rl\rr}$ and $\delta Q_{\rl\rr}$ generically have additional degrees of freedom, and the limit is not expected to depend on the initial density matrix for these degrees of freedom.

The existence of a non-equilibrium steady state is not {\em a priori} immediate, and has been studied in various cases. In the context of quantum dots, there is a finite-dimensional small quantum system in the above description (the dot), and the reservoirs are one-dimensional massless free fermions (the $s$-waves of the electronic leads). The proof of the existence of the steady state limit exists in the resonant-level model \cite{berjmp}, and to all order of a perturbative expansion in the Kondo dot \cite{doyand}. In situations where $H$ is homogeneous, there are proofs of the existence of the steady state limits for chains of free fermionic degrees of freedom \cite{tas}, for the Ising, XX and XY quantum chains \cite{ah,og,aschp}, for any low-energy one-dimensional quantum critical chain within the formalism of CFT \cite{bd1,bd2}, and for free field theory of any dimensionality \cite{collura2,doyonKG}. In the following we assume the steady-state limit to exist for the observables considered, and from now on the limit $L\to\infty$ is understood.

\section{Ballistic transport} \label{sectbal}

After a long time $t$ following connection time, one can roughly divide the full  system into five regions, see Fig. \ref{quench}. Due to the finite Lieb-Robinson velocity \cite{lr} (recall that it is normalized to unity), there is a ``light-cone'' effect: there are two regions $x^1<-t$ and $x^1>t$ where the reservoirs are locally thermal with ``exponential'' precision (the asymptotic baths). There are two transition regions that interpolate between the asymptotic baths and the area near the system, $x^1\in[-t,-\ell]$ and $x^1\in[\ell,t]$ with $0<\ell\ll t$, and the final region $x^1\in[-\ell,\ell]$ is where the steady-state limit exists (the stationary region). The value of $\ell$ is to a large extent arbitrary: it is only to be kept finite in the steady-state limit.

Asymptotic baths become further away from the connection interface with time. Therefore one expects gradients of densities in the transition regions to become smaller. Hence, diffusive transport, controlled by gradients (e.g.~Fourier's law), is absent in the steady state. That is, quantities with nonzero flows in the non-equilibrium steady state are only those that are subject to ballistic transport.
This suggests the following definition. Let $Q$ be a local conserved charge whose transport we wish to study, $Q = \int \dd^d x\,\qq(x)$, $[H,Q]=0$. By locality and conservation, there is a current $\ve\jj$ satisfying
\beq\label{qj}
	\p_t \qq + \nabla \cdot\ve\jj=0
\eeq
(and recall that $\jj:=\ve\jj^1$ is the longitudinal component). We say that the quantity $Q$ is subject to non-equilibrium ballistic transport under the dynamics $H$ and the given initial imbalance, if the steady state $\sta{\cdot}$ is homogeneous (does not spontaneously break the translation symmetry of $H$) and the steady-state average $\sta{\jj}:=\sta{\jj(x)}$ is nonzero. In the cases mentioned above where a non-equilibrium steady state has been shown to exist, one can verify explicitly that the homogeneous reservoirs involved indeed admit ballistic transport of the quantity studied.

The above definition is a far-from-equilibrium concept. Ballistic transport is often discussed near equilibrium, in terms of the Drude peak \cite{maz,zot}, related to the current two-point function. From the Drude peak analysis one concludes that if the total current $J = \int\dd^d x\,\jj(x)$ is conserved, then the linear-response conductivity is nonzero (more generally, by Mazur's inequality \cite{maz}, one only requires the presence of conserved quantities with which $J$ has nonzero ``overlap''). If $\jj(x)$ is a conserved local density as in \eqref{JI}, then indeed $J$ is a conserved quantity. However, far from equilibrium, the relation between these concepts and ballistic transport is less clear. Instead, the statement \eqref{jineq} about the value of $\sta\jj$ provides an answer as per the above non-equilibrium definition. We now proceed to show \eqref{jineq}.

\section{Regularity condition and proof of Inequality \eqref{jineq}} \label{sectproof}
Let $\jj(x)$ be a conserved local density, so that \eqref{JI} holds.
We impose the following {\em regularity} condition on $\uu$: there exists $C(\ell)$ with $\lim_{\ell\to\infty} C(\ell) =: C_\infty \geq 0$, such that for all $x^1\neq0$ and all $\alpha<1$ near enough to 1,
\beq\label{assineq}
	-\frc1{x^1} \int_{\alpha|x^1|}^\infty \dd t\, \big(\ini{\uu(x,t)}-\sta\uu\big)
	\geq C(|x^1|).
\eeq

It is natural to assume that $\gibl\uu\geq\gibr\uu$, if the initial reservoirs are thermalized as to produce a positive current from the left to the right. Then the regularity condition is interpreted as the fact that the average of $\uu$ in the transient region does not pass its stationary value by too large an amount or for too long a time. That is, for fixed $x^1<0$ ($x^1>0$) and large $|x^1|$, the value of $\ini{\uu(x,t)}$, as $t$ increases, does not go too far, or for too long a time, below (above) its stationary value $\sta{\uu}$. This condition is satisfied, for instance, if $\ini{\uu(x,t)}$ is monotonic in $t$ within the transient regions, at least far enough from the connection interface and on large scales. It may be broken, for instance, if $\ini{\uu(x,t)}$, say for $x^1<0$, takes a small enough value, smaller than $\sta\uu$, over a period of time that grows like $|x^1|$.

Inequality \eqref{jineq} is shown under the regularity condition as follows. First note that in dimensions $d>1$, one may consider averages over the large transverse hyperarea, $\h\jj(x^1),\h\uu(x^1) := \lim_{V_\perp\to\infty} V_\perp^{-1} \int_{V_\perp} \dd^{d-1}x^\perp \jj(x), \uu(x)$. The $d$-dimensional equation \eqref{JI} implies that these averaged quantities satisfy the one-dimensional version of \eqref{JI}. Using transverse homogeneity of the full time-evolved state, one may scale out the transverse hyperarea $V_\perp$ when evaluating quantum averages. Hence the problem reduces to one dimension.

Let us then consider $d=1$. By the conservation equation \eqref{JI} we have, for any $t,\ell>0$ (see Fig. \ref{quench}),
\beq
	\int_{-\ell}^\ell \dd x\,\big(\jj(x,t) - \jj(x,0)\big) =
	\int_0^t \dd s\,\big(\uu(-\ell,s) - \uu(\ell,s)\big).
\eeq
When evaluated in the initial state $\ini{\cdot}$, we may use $\ini{\jj(x,0)}=0$.
Taking $t\to\infty$ with fixed $\ell$ and $\alpha<1$,
\beqa
	\sta{\jj} &=& 
	\frc1{2\ell}\int_0^{\alpha\ell} \dd s\, \big(\ini{\uu(-\ell,s)} - \ini{\uu(\ell,s)}\big) + \n
	&+& 
	\frc1{2\ell}\int_{\alpha\ell}^\infty\dd s\big(\ini{\uu(-\ell,s)} - \ini{\uu(\ell,s)}\big). \label{meq}
\eeqa

We evaluate the first line on the right-hand side of \eqref{meq} in the limit $\ell\to\infty$ using two ingredients.

The first ingredient is the Lieb-Robinson bound \cite{lr}, which leads to the following (see Appendix \ref{appLR}): for every $a>0$ small enough there exists $v_a$, such that for every bounded operator $\ww$ and $\t \ww$ and every $t>0$, the inequality $||[\ww(t),\t\ww(0)]||\leq A e^{-a(D(\ww,\t \ww)-v_a t)}$ holds for some $A>0$. Here $||\cdot||$ is the operator norm and $D(\ww,\t\ww)$ is the distance between the supports of $\ww$ and $\t\ww$. The Lieb-Robinson velocity is the infimum $v = {\rm inf}(v_a:a>0)$ (which we have normalized to $v=1$). The Lieb-Robinson bound implies that we may approximate $\ww(t)$ by an operator $[\ww(t)]_r$ supported on a neighborhood extending a distance $r$ from the support of $\ww$, such that in any state $\gen{\cdot}$,
\beq\label{LR}
	|\gen{\ww(t)}-\gen{[\ww(t)]_r}| \leq \t A e^{-a(r-v_at)}.
\eeq
The second ingredient is the fact that bulk averages are obtained in the limit where operators are far from any boundary:
\beq\label{boundbulk}
	\lim_{D(\ww,0)\to\infty} \ini{\ww} = \gen{\ww}_{\rl,\rr}\quad
	\mbox{($\ww$ supported on the left/right)}.
\eeq

With $\ell' = \alpha\ell$ and $\ell''=\sqrt{\alpha}\ell$, we now bound the integral $\int_0^{\ell'} ds\,\big(\ini{\uu(\ell,s)}-\gen{\uu}_\rr\big) = \int_0^{\ell'} ds\,\big(\ini{\uu(\ell,s)}-\gen{\uu(\ell,s)}_\rr\big)$ as follows:
\beqa
	\lefteqn{\lt|\int_0^{\ell'} ds\,\big(\ini{\uu(\ell,s)}-\gen{\uu}_\rr\big)\rt| } \qquad\qquad&& \n &\leq&
	\int_0^{\ell'} \big| \ini{\uu(\ell,s)}-\ini{[\uu(\ell,s)]_{\ell''}}\big| +
	\n
	&+& 
	\int_0^{\ell'} \big| \gen{\uu(\ell,s)}_\rr-\gen{[\uu(\ell,s)]_{\ell''}}_\rr\big| + \n
	&+& 
	\int_0^{\ell'} \big| \ini{[\uu(\ell,s)]_{\ell''}}-\gen{[\uu(\ell,s)]_{\ell''}}_\rr\big|. \no
\eeqa
Using \eqref{LR} with  $v_a = 1/\sqrt{\alpha}>1$, the first two lines on the right-hand side give $O(1)$ as $\ell\to\infty$. Using \eqref{boundbulk} and $\alpha<1$, the last line gives $o(\ell)$. A similar calculation holds for the integral of $\ini{\uu(-\ell,s)} - \gen{\uu}_\rl$. Hence we find
\beq
	\lim_{\ell\to\infty} \frc1{2\ell}\int_0^{\alpha\ell} \dd s\, \big(\ini{\uu(-\ell,s)} - \ini{\uu(\ell,s)}\big) 
	= \alpha\frc{\gen{\uu}_\rl - \gen{\uu}_\rr}2 \label{der1}
\eeq
and we may take the limit $\alpha\to1^-$.

The second line on the right-hand side of \eqref{meq} is evaluated in the limit $\ell\to\infty$ using \eqref{assineq}:
\beqa
	\lim_{\ell\to\infty} \frc1{2\ell}\int_{\alpha\ell}^\infty\dd s\big(\ini{\uu(-\ell,s)} - \ini{\uu(\ell,s)}\big)
	&\geq&
	\lim_{\ell\to\infty} C(\ell)\n
	&\geq & 0. \label{der2}
\eeqa
Combining  \eqref{meq}, \eqref{der1} and \eqref{der2} we obtain \eqref{jineq}.

A small modification of the above proof gives rise to the inequalities $v_{\rl,\rr}<1$ for the left and right transient velocities \eqref{vlr}.

\section{Non-linear and generalized sound velocities}\label{sectsound}

The inequality \eqref{jineq} naturally suggests the definition \eqref{veff} of the transient velocity $\vtr$, as it bounds $\vtr$ by the Lieb-Robinson velocity, and the stronger relations suggest $v_{\rl,\rr}$. From the above proof and the examples below, one can interpret $\vtr$, and in more details $v_{\rl,\rr}$, as effective velocities characterizing the transient regions. We now argue that, in non-integrable models, these velocities have the interpretation as generalizations, beyond the linear-response regime, of the sound velocity $v_{\rm s}$. Our assumptions are that near equilibrium, and after large enough times, there is local Gibbs thermalization, and that an equation of state relates Gibbs averages $\gen\cdot$, of the form $\gen{\uu} = F(\gen{\qq})$, with $F$ having nonnegative first derivative. We will show \eqref{limv} under these assumptions, where
\beq\label{vs}
	v_{\rm s}=\sqrt{F'(\gen\qq)}
\eeq
is the velocity of small planar waves about the Gibbs state, describing the propagation of $\qq$ in the longitudinal direction $x^1$, and emerging from the wave equations \eqref{JI} and \eqref{qj} under the equation of state.

The above assumptions are expected to hold in interacting, non-integrable models. In integrable models, the assumption of local Gibbs thermalization fails. One expects rather local ``generalized Gibbs ensembles'' to occur, because of the presence of infinitely many conservation laws \cite{rigol}. In this case, the derivation below does not apply, and the transient velocities do not specialize to the sound velocity associated to the Gibbs equation of state, as is confirmed by examples in the next section. Surprisingly, this is true even though initial states on the left and right are Gibbs states, and are taken near to each other. The left-hand side of the limit \eqref{limv} is then naturally interpreted as ``generalized sound velocities'' in integrable models, quantities that are consequences of near-equilibrium generalized thermalization.

We denote by $\gen{\cdot}$ the equilibrium state occurring in the limit taken in \eqref{limv}. Consider the quantities $\h\qq$, $\h\jj$ and $\h\uu$, which are densities averaged over the transverse hypersurface as in the previous section, $\h\qq(x^1), \h\jj(x^1),\h\uu(x^1) = \lim_{V_\perp\to\infty} V_\perp^{-1} \int_{V_\perp}\dd^{d-1}x^\perp \,\qq(x),\jj(x),\uu(x)$. Again as in the previous section, Equations \eqref{JI} and \eqref{qj} imply their one-dimensional versions on these averaged quantities. Using homogeneity in the transverse direction for quantum averages, it is then sufficient to set $d=1$.

In the linear response regime, the averages are described by small variations around the equilibrium state, $\ini{\qq(x,t)} \approx \gen{\qq} + \delta\qq(x,t)$, $\ini{\jj(x,t)} \approx \delta\jj(x,t)$ and $\ini{\uu(x,t)} \approx \gen{\uu} + \delta\uu(x,t)$. Within linear response, we may further assume that at every space-time point, the system is approximately at equilibrium. Using the equation of state, this implies that $\ini{\uu(x,t)} \approx F(\ini{\qq(x,t)})$, whence $\delta\uu(x,t) = F'(\gen{\qq})\,\delta\qq(x,t)$. Putting these into \eqref{JI} and \eqref{qj}, we find the wave equation
\beq
	\p_t^2 \delta\qq(x,t) = F'(\gen{\qq})\,\p_x^2 \delta\qq(x,t).
\eeq
Hence, we can identify the sound velocity as \eqref{vs}. The wave equation implies that
\beq\label{dq}
	\delta\qq(x,t) = f(x-v_{\rm s}t) + g(x+v_{\rm s}t).
\eeq
Using \eqref{JI} and \eqref{qj}, we then find the expression
\beq\label{dj}
	\delta \jj(x,t) = v_{\rm s} (f(x-v_{\rm s}t)-g(x+v_{\rm s}t)).
\eeq

From these results, we may infer the linear-response variations of the stationary current with respect to variations of left and right reservoirs' states near the common equilibrium state. From \eqref{dq}, reservoirs' variations are given by $\delta\gen{\qq}_{\rl,\rr} = f(\mp\infty) + g(\mp\infty)$, and using \eqref{dj} along with the fact that the current is zero in the reservoirs, we have $f(\pm\infty) = g(\pm\infty)$. Further, from \eqref{dj}, we find the small steady-state current to be $\delta\sta{\jj} = v_{\rm s}( f(-\infty)-g(\infty))$, giving the relation
\beq
	\delta\sta{\jj}  = \frc{v_{\rm s} (\delta\gen{\qq}_\rl - \delta\gen{\qq}_\rr)}2 = \frc{\delta\gen{\uu}_\rl - \delta\gen{\uu}_\rr}{2 v_{\rm s}}.
\eeq
This gives \eqref{limv} for $\vtr$. Using further $\delta \gen{\qq}_\rs = f(-\infty)+g(\infty)$ gives \eqref{limv} for $v_{\rl,\rr}$.

\section{Models analysis}\label{sectmod}

We may test inequality \eqref{jineq} in explicit models where exact results are available. Details of these model calculations are presented in Appendix \ref{appmod}; here we discuss the results.

The inequality is {\em saturated} in low-energy one-dimensional quantum critical systems with unit dynamical exponent, for energy and charge transport. In both cases CFT chiral separation $\qq(x,t) = \qq_+(x-t) + \qq_-(x+t)$ implies that $\uu = \qq$ (for energy transport, this is tracelessness of the stress-energy tensor), and, as is explained in \cite{bd1,bd2}, it also implies that $\sta\jj = (\gen{\qq}_{\rl}-\gen{\qq}_\rr)/2$. Saturation is understood by the fact that under time evolution, thanks to chiral separation, sharp ``shock waves'' appear at the Lieb-Robinson speed, separating the exact steady state from exact reservoirs (the transient region in Fig. \ref{quench} is of microscopic width). Hence there is no correction coming from the transient region: the function $C(\ell)$ vanishes and $\vtr=v_\rl =v_\rr=1$.

The inequality, however, appears to be {\em strict} in many situations away from one-dimensional criticality.

In CFT in any dimension, the steady state for energy transport, according to \cite{bhaseen1}, is a boosted thermal state. In $d>1$ dimensions, expressions for the current $\sta{\jj} = \sta{\TT^{01}}$
and pressures $\gen{\uu}_{\rl,\rr} = \gen{\TT^{11}}_{\rl,\rr}$ were proposed from QFT methods, gauge-gravity duality and hydrodynamics ideas in  \cite{bhaseen1}, and from hydrodynamics considerations in \cite{cky}. Employing the results and formulae of \cite{bhaseen1}, the transient velocity is
\beq
	\vtr = \frc{d+1}{2d}\frc{t_\rl+t_\rr}{\sqrt{t_\rl+dt_\rr}\sqrt{t_\rl+d^{-1}t_\rr}}<1\quad\mbox{(CFT)}
\eeq
where $t_{\rl,\rr} = T_{\rl,\rr}^{(d+1)/2}$. The inequality \eqref{jineq} is strict.
The dynamics is expected to gives rise to thin (of extent $o(t)$) transient regions centered on world-surfaces at speeds $v_{\rl,\rr} = \sqrt{t_{\rr,\rl}+d^{-1} t_{\rl,\rr} }/\sqrt{t_{\rr,\rl}+ dt_{\rl,\rr} }<1$, which generalize the shock waves of the one-dimensional case and indeed satisfy $\vtr = (v_\rl + v_\rr)/2$. As noted in \cite{bhaseen1}, both $v_{\rl,\rr}$ tend to the conformal relativistic sound velocity $v_{\rm s}=1/\sqrt{d}$ at equilibrium, in agreement with the general result of Section \ref{sectsound}.

In free critical models, the steady state is instead described by independently thermalized Fock modes. For energy transport in the massless Klein-Gordon (KG) theory, using results of \cite{doyonKG}, we find
\beq\label{vtrkg}
	\vtr = \frc{\sqrt{\pi}\Gamma((d+1)/2)}{d\Gamma(d/2)}\leq 1
	\quad\mbox{(masless KG)}
\eeq
with equality only at $d=1$. In this case the transient regions are large and in the semiclassical (large-scale) approximation the pressures are monotonic, as independent excitations of various group velocities slowly build the steady state region, thus \eqref{assineq} is satisfied. Note that, in contrast with the interacting CFT case, the transient velocity is temperature-independent. Further, it is {\em greater} than the conformal relativistic sound velocity. This is a consequence of integrability of the KG theory, and \eqref{vtrkg} is a generalized sound velocity.

The inequality also appears to be strict away from criticality. In the Klein-Gordon model with a mass $m$, one multiplies the above $\vtr$  \eqref{vtrkg} by the (temperature-dependent) factor \cite{doyonKG}
\[
	\frc{\int_0^\infty \dd p\,p^{d+1}/E_p\,(b_{\rl}(p)-b_{\rr}(p))}{
	\int_0^\infty \dd p\,p^d\,(b_{\rl}(p)-b_{\rr}(p))}\leq1
\]
where $b_{\rl,\rr}(p) = (e^{\beta_{\rl,\rr} E_p} -1)^{-1}$ are bosonic thermal distributions and $E_p = \sqrt{p^2+m^2}$ is the relativistic energy at momentum $p$. This equals 1 only at $m=0$. The picture is again of large transient regions and monotonic pressure. The generalized sound velocity obtained at $\beta_\rl=\beta_\rr=:\beta$, in this case, is temperature dependent, and equals the quantity \eqref{vtrkg} times
\[
	\frc{\int_0^\infty \dd p\,p^{d+1}/\sinh^2(\beta E_p/2)}{
	\int_0^\infty \dd p\,p^dE_p/\sinh^2(\beta E_p/2)}.
\]

All examples above are within field theory. There are fewer exact results away from universality, however we may analyze the transverse-field Ising model, which has a free-fermion representation. The non-equilibrium current was studied in \cite{ah,aschp,delucaising}, and the equilibrium pressures may be evaluated by using the standard exact solution. Normalizing the Hamiltonian so that the Lieb-Robinson velocity is unity (see Appendix \ref{appmod}), we find
\beq
	\vtr = \frc{\int_0^\pi \dd\theta \, h \sin^2\theta \,/ \ep(\theta)\; (f_\rl(\theta)-f_\rr(\theta))}{
	\int_{0}^\pi \dd\theta \,\sin\theta\,(f_\rl(\theta)-f_\rr(\theta))} \leq 1
	\quad\mbox{(Ising)}
\eeq
where $h\geq 1$ is the transverse magnetic field, $f_{\rl,\rr}(\theta) = (e^{\beta_{\rl,\rr} \ep(\theta)} +1)^{-1}$ are fermionic thermal distributions and $\ep(\theta) = \sqrt{h^2+1-2h\cos\theta}$ is the single-particle energy at wave number $\theta$.

For massive integrable models of QFT, the equilibrium thermodynamic Bethe ansatz (TBA) \cite{tba1} has been conjecturally generalized to non-equilibrium steady states \cite{doyonint} following general ideas of \cite{bd1}, giving integral equations for the non-equilibrium energy current. Similar ideas have been used for the XXZ chain \cite{DeLucaVitiXXZ}. As was pointed out in \cite{DeLucaVitiXXZ}, such expressions should only be expected to be approximations within certain regimes of validity, albeit more accurate than linear response. A numerical analysis of the self-dual sinh-Gordon and the roaming-trajectory models shows that \eqref{jineq} is indeed satisfied only for small enough temperatures. We hope to come back to these issues in a future work.

\section{Relativistic invariance} \label{sectrel}

The bound in \eqref{jineq} involves equilibrium values of $\uu$, but it may be difficult in specific situations to argue that $\gibl\uu>\gibr\uu$ or to determine the difference more precisely. In relativistic models certain response functions satisfy additional relations, which allow us to obtain explicitly positive bounds. In particular, for pure energy transport between reservoirs at inverse temperatures $\beta_\rl$ and $\beta_\rr$, the right-hand side \eqref{jineq} (with $\jj=\TT^{01}$ and $\uu=\TT^{11}$) can be re-written as an explicitly positive quantity using these thermodynamic relations:
\beqa\label{relener}
	\gibl{\TT^{11}} - \gibr{\TT^{11}} &=& 
	\int_{\beta_\rl}^{\beta_\rr} \dd\beta
	\int \dd^dx\,\gen{{\TT^{01}}(x){\TT^{11}}(0)}_{\beta} \n &=& 
	\int_{\beta_\rl}^{\beta_\rr} \frc{\dd\beta}{\beta}
	\big(\gen{\TT^{00}}_\beta+\gen{\TT^{11}}_\beta\big)
\eeqa
where $\gen{\cdot}_\beta$ is a thermal state at inverse temperature $\beta$.  One can also simplify the right-hand side of \eqref{c2} by applying $-\p/\p\beta_\rl +\p/\p\beta_\rr$ on \eqref{relener} (making it explicitly positive).

Consider a density matrix of the form $e^{-\beta \lt(H-\sum_j \mu^{(j)} Q^{(j)}\rt) + \nu P}$, where  $\nu$ is associated to the momentum $P=\int \dd^{d} x\,\TT^{01}(x)$. Define $\t\mu^{(j)}:=\beta \mu^{(j)}$. Denoting by $\gen{\cdot}_{\beta,\nu,\t\mu^{(j)}}$ the corresponding state, if current conservation \eqref{JI} holds, then the following thermodynamic relation is valid, proved in Appendix \ref{proofrel}:
\beq\label{addr}
	-\frc{\p}{\p\beta} \gen{\uu}_{\beta,\nu,\t\mu^{(j)}} =
	\frc{\p}{\p\nu} \gen{\jj}_{\beta,\nu,\t\mu^{(j)}}.
\eeq

Suppose that the left and right reservoirs are $\gibl{\cdot} = \gen{\cdot}_{\beta_\rl,0,\t\mu^{(j)}}$ and $\gibr{\cdot} = \gen{\cdot}_{\beta_\rr,0,\t\mu^{(j)}}$ with $\beta_\rl<\beta_\rr$. A natural interpretation of the right-hand side of \eqref{addr},
at $\nu=0$, is as the linear response of the current to a small boost, which should be positive (thus implying a positive bound in \eqref{jineq}). In fact, integrating \eqref{addr} over $\beta$, we find from \eqref{jineq}
\beq\label{jineqrel}
	\sta{\jj} \geq \frc12 \int_{\beta_\rl}^{\beta_\rr} \dd\beta
	\int \dd^d x\,\gen{\TT^{01}(x)\jj(0)}_{\beta,\nu=0,\t\mu^{(j)}}
\eeq
(where imaginary time-ordering is implied). Specializing to the energy current $\jj= \TT^{01}$, the two-point function is explicitly positive by reflection positivity, and we have the first line of \eqref{relener}. In pure thermal states $\gen{\cdot}_\beta = \gen{\cdot}_{\beta,0,0}$, the pressure is related to the specific free energy $f$ as $\gen{\TT^{11}}_\beta = -f$ and the energy density as $\gen{\TT^{00}}_\beta = \p(\beta f)/\p\beta$. Using \eqref{addr} we then obtain $\beta \int \dd^d x\,\gen{\TT^{01}(x)\TT^{01}(0)}_{\beta} = \gen{\TT^{00}}_\beta + \gen{\TT^{11}}_\beta$, giving the second line of \eqref{relener}.

\section{An inequality for the near-equilibrium noise}\label{sectnoise}

For simplicity we consider pure energy transport between thermal reservoirs, $\gen{\cdot}_{\rl,\rr} = \gen{\cdot}_{\beta_{\rl,\rr}}$ with $\beta_\rl<\beta_\rr$. Let us denote $\sta{\jj}=\gen{\jj}_{\rs;\beta_\rl,\beta_\rr}$. The integral form of \eqref{jineq} is
\beq\label{integr}
	\int_{\beta_\rl}^{\beta_\rr} \dd \beta\,
\lt[-\frc{\p}{\p\beta} \gen{\jj}_{\rs;\beta,\beta_\rr} + \frc{\p}{\p\beta} \frc{\gen{\uu}_\beta}2\rt]\geq 0.
\eeq
A similar form holds by varying $\beta_\rr$ instead of $\beta_\rl$. Hence in order to verify \eqref{jineq} it is sufficient to have either of the following two inequalities:
\beq\label{jineqc}
	\mp\frc{\p}{\p\beta_{\rl,\rr}} \gen{\jj}_{\rs;\beta_\rl,\beta_\rr} \geq
	- \frc{\p}{\p\beta_{\rl,\rr}}\, \frc{\gen{\uu}_{\beta_{\rl,\rr}}}2.
\eeq
Note that one implies the other if the current is anti-symmetric under $\beta_\rl \leftrightarrow \beta_\rr$. These inequalities are not necessary consequences of (\ref{jineq}). However, (\ref{integr}) must hold for every $\beta_\rl$, hence the maximum of the integrand is positive on every interval $[\beta_\rl,\beta_\rr]$ (fixed $\beta_\rr$). By continuity, the inequalities \eqref{jineqc} hold for all $|\beta_\rl-\beta_\rr|$ small enough (i.e.~near enough to equilibrium). A similar argument can be used to show \eqref{Gineq}.

An interesting consequence is an inequality for the energy transfer noise $c_2:=\int \dd t\,\dd^{d-1}x^\perp\,\lt(\sta{\jj(x^\perp,t) \jj(0)} - \sta{\jj}^2\rt)$. According to the extended fluctuation relations (EFR) \cite{doyontrans}, it is obtained by differentiating the current, $c_2 = (-\p/\p\beta_\rl + \p/\p\beta_\rr) \gen{\jj}_{\rs;\beta_\rl,\beta_\rr}$. The EFR have been shown in integrable models \cite{doyontrans} and at criticality \cite{bd1,bd2,bhaseen1} in any dimensionality, and can be derived from PT symmetry of the non-equilibrium steady-state \cite{bhaseen1}. We expect the EFR to be valid for ballistic transport in quite generally. For instance the relations can be argued to hold away from integrability for PT symmetric systems, following the insight of \cite{bhaseen1}, as follows: in the partitioning approach, one expects local ``thermalization'' to occur as time evolves, and the steady state to be described by a density matrix involving the local conserved charges available; generically this leads to $e^{-\beta H + \nu \int \dd^d x\,\jj(x)}$, which is expected to be PT symmetric in most cases.

Assuming the EFR and using \eqref{jineqc}, in a nonvanishing region near enough to equilibrium we then have \eqref{c2}.

Surprisingly, it turns out that for all models where we have verified (\ref{jineq}), the strong inequalities (\ref{jineqc}) are satisfied {\em for all} $\beta_\rl$ and $\beta_\rr$ (and saturated at one-dimensional criticality). We do not know yet  of a general proof or of natural dynamical conditions leading to this stronger statement.

\section{Conclusion}\label{sectcon}

We have obtained inequalities bounding, under certain conditions, the non-equilibrium steady-state current, and the noise near enough to equilibrium, in the partitioning approach for general local quantum many-body systems. These inequalities in particular show the presence of ballistic transport under these conditions. This generalizes to far-from-equilibrium setups the well-known conclusions, near equilibrium, from an analysis of the Drude peak, although it is based on somewhat different ideas. We have verified the inequalities in various examples, including integrable and non-integrable field theory models and the Ising quantum chain.

In the higher-dimensional setup, we assumed homogeneity of the state in the transverse direction during the whole time evolution. Perturbations in the transverse directions will affect this assumption, as turbulent effects may develop. In the above derivations, quantities involved will be transversal averages instead of proper densities. In particular, the statement of the regularity condition will be thus modified. However, with this precision, our main conclusions remain.

An interesting physical principle emerging from this analysis is that of non-linear sound velocities. The inequalities are the statements that non-linear sound velosities are bounded from above by the Lieb-Robinson velocity. Near equilibrium they specialize to the sound velocity in non-integrable systems, but interestingly they do not in integrable models. The latter is an effect of generalized thermalization due to the presence of infinitely many conservation laws. It would be very interesting to investigate further near-equilibrium wave propagation in integrable models by developing generalized Gibbs thermalization ideas.

We observed that the inequality is broken by recent TBA formulations of non-equilibrium steady states in certain regimes of temperatures. It will be important to clarify both the physics of non-equilibrium steady state formation in interacting integrable models, and the range of validity of TBA formulations.

Recall that the usual near-equilibrium Drude peak analysis also provides information about ballistic transport of quantities that {\em do not} correspond to wave equations (where the current is not the density of a conservation law \eqref{JI}), using the notion of overlap with conserved quantities. It would be very interesting to likewise generalize the present far-from-equilibrium analysis to such quantities.

Finally, it remains to elucidate the conditions under which the stronger inequality \eqref{jineqc} should hold.

{\bf Acknowledgments.} I warmly thank Denis Bernard, Joe Bhaseen, \'Edouard Boulat and Olalla Castro-Alvaredo for encouragements and discussions, and in particular Olalla Castro-Alvaredo for confirming the numerical TBA results. I am also  indebted to my other collaborators on this general subject, Koenraad Schalm and Andrew Lucas. Finally I am grateful to Universit\'e Paris Diderot, where part of this work was done, for financial support through a visiting professorship.

\appendix

\section{Lieb-Robinson bound} \label{appLR}

The original expression \cite{lr} of the Lieb-Robinson bound is as follows: there exists a $v>0$ and a strictly positive increasing function $\mu(w)$ such that for every $\ww,\t\ww$ and every $w>v$,
\beq
	\lim_{t\to\infty\atop D(\ww,\t\ww)\geq w t} e^{\mu(w) t} ||
	[\ww(t),\t\ww(0)]|| = 0.
\eeq
This implies $||[\ww(t),\t\ww(0)]|| \leq A e^{-\mu(w)t}$ for every $\ww,\t\ww$, $t>0$ and $w>v$ with $D(\ww,\t\ww)\geq wt$ (where $A$ depends on $\ww,\t\ww$). Denote $D(\ww,\t\ww) = w' t$. Then $D(\ww,\t\ww) -wt = (w'-w)t$ and we have, using the inequality for $w$ replaced by $w'$,
\beqa
	||[\ww(t),\t\ww(0)]|| &\leq& A e^{-\frc{\mu(w')}{w'-w}(D(\ww,\t\ww)-wt)}
	\n &\leq&
	A e^{-{\rm inf}\lt(\frc{\mu(w')}{w'-w}:w'>w\rt)(D(\ww,\t\ww)-wt)}.
\eeqa
According to the proof in \cite{lr} (p. 256), $\mu(w)\propto w$ for $w$ large, hence the infimum is nonzero. We choose $a \leq {\rm inf}\lt(\frc{\mu(w')}{w'-w}:w'>w\rt)$ and $v_a=w$ to obtain the formulation presented in the main text; the condition $D(\ww,\t\ww)\geq wt$ is not needed by boundedness of the operators.

\section{Models calculations} \label{appmod}

\subsection{Conformal field theory}

Non-equilibrium steady states in the present setup have been studied one-dimensional CFT using chiral factorization in \cite{bd1,bd2}, and in higher-dimensional interacting CFT using QFT methods, gauge-gravity duality and hydrodynamics arguments in \cite{bhaseen1}, and using hydrodynamics arguments in \cite{cky}. The one-dimensional case agrees with the limit $d\to1$ of the higher-dimensional case, hence we will present the general-$d$ result. It was found that the steady states are described by boosted thermal states. Symmetries of the problem and tracelessness of the stress-energy tensor then imply that $\sta{T^{\mu\nu}} = a_d\,T^{d+1}(\eta^{\mu\nu}+(d+1)u^\mu u^\nu)$ where $\eta^{\mu\nu} = {\rm diag}(-1,1,\ldots,1)^{\mu\nu}$ and $u^\mu = (\cosh\theta,\sinh\theta,0,\ldots,0)^\mu$. Here $T$ is the rest-frame temperature and $\theta$ the boost rapidity. The values of $T$ and $\theta$  in terms of $T_\rl$ and $T_\rr$, as evaluated in \cite{bd1} for $d=1$ and proposed in \cite{bhaseen1} for $d>1$, give
\beq
	\sta{\TT^{01}} =
	d \,a_d/(d+1)\, (t_\rl-t_\rr) \sqrt{t_\rl+dt_\rr}\sqrt{t_\rl+d^{-1}t_\rr}
\eeq
where $t_{\rl,\rr} = T_{\rl,\rr}^{(d+1)/2}$. The constant $a_d$ is model-dependent, and specializes to $a_1=c/6$ where $c$ is the central charge. At equilibrium (i.e. setting $\theta=0$), we have $\gen{T^{\mu\nu}} = a_d T^{d+1}{\rm diag}(d,1,\ldots,1)^{\mu\nu}$. Hence the pressures are $\gen{\TT^{11}}_{\rl,\rr} = a_d\, t_{\rl,\rr}^2$. This leads to the  transient velocity $\vtr$ shown in the main text, which fufills \eqref{jineq}.

Further, differentiating we obtain
\beqa
	\lefteqn{-\frc{\p}{\p\beta_\rl} \gen{\TT^{01}}_{\rs;\beta_\rl,\beta_\rr}} \hspace{1cm}
	&& \n &=& \frc{d\,a_d}{2}t_\rl^{\frc{d+3}{d+1}} \times \n && \times\;
	\lt(1+\frc{t_\rl-t_\rr}{2(t_\rl+d t_\rr)} + \frc{t_\rl-t_\rr}{2(t_\rl+d^{-1} t_\rr)}\rt)\times \n&&\times\; \sqrt{t_\rl+dt_\rr}\sqrt{t_\rl + d^{-1}t_\rr}
\eeqa
and
\beq
	-\frc12 \frc{\p}{\p\beta_\rl} \gen{\TT^{11}}_{\beta_\rl} = \frc{(d+1)\,a_d}2
	t_\rl^{\frc{d+3}{d+1}+1}.
\eeq
An analysis of these expression, using $(t_\rl+dt_\rr)(t_\rl + d^{-1}t_\rr) \geq (t_\rl + t_\rr)^2$ and $\sqrt{t_\rl+dt_\rr}/\sqrt{t_\rl + d^{-1}t_\rr} + \sqrt{t_\rl+d^{-1}t_\rr}/\sqrt{t_\rl + dt_\rr} \geq 2$, shows that the first of \eqref{jineqc} indeed holds (hence the second holds by anti-symmetry of the current).

\subsection{Klein-Gordon model}

In free (quadratic) models, the general statement is that a non-equilibrium density matrix $\rho_\rs$ reproducing $\sta{\cdot}$ can be written in terms of free modes, where those with positive (negative) momenta are thermalized with $\beta_\rl$ ($\beta_\rr)$. In the case of the Klein-Gordon theory this was shown in \cite{doyonKG}, where various observables including the energy current and its fluctuations were studied within the partitioning approach. It was found that
\beq
	\rho_\rs = e^{- \int Dp\, W(p) A^\dag(p)A(p)}
\eeq
where $Dp = d^dp/((2\pi)^d 2E_p)$ is the relativistically invariant measure (we recall that $E_p = \sqrt{p^2+m^2}$), $W(p) = (\beta_\rl \Theta(p^1) + \beta_\rr \Theta(-p^1))\,E_p$ describes the independent thermalization of right-moving and left-moving modes ($\Theta$ is the step function), and $A(p)$, $A^\dag(p)$ are annihilation and creation operators at momentum $p$ with relativistic normalization $[A(p),A^\dag(p')] = (2\pi)^d 2E_p\,\delta^{(d)}(p-p')$. Following textbook material, the stress-energy tensor can be written explicitly in terms of these operators, and one finds \cite{doyonKG}
\beqa
	\sta{\TT^{01}} &=& d\,\Gamma(d/2)\zeta(d+1)/(2\pi^{\frc d2+1})\,(r_\rl^2 - r_\rr^2),\n
	\gen{\TT^{11}}_{\rl,\rr} &=& \Gamma((d+1)/2)\zeta(d+1)/(\pi^{\frc{d+1}2})\,s_{\rl,\rr}^2
\eeqa
where $r_{\rl,\rr}=s_{\rl,\rr}= T_{\rl,\rr}^{(d+1)/2}$ if the mass $m$ is zero, and
\beqa
	r_{\rl,\rr}^2 &=& \frc{1}{d! \zeta(d+1)} \int_0^\infty \dd p\,\frc{p^d}{e^{\beta_{\rl,\rr} E_p}-1}\n
	s_{\rl,\rr}^2 &=& \frc{1}{d! \zeta(d+1)} \int_0^\infty \dd p\,\frc{p^{d+1}}{E_p\, (e^{\beta_{\rl,\rr} E_p}-1)}
\eeqa
if $m\neq 0$. Here $\zeta(z)$ is Riemann's zeta function. These lead to the expressions of $\vtr$ written in the main text. Further, using the facts that $d\,\Gamma(d/2)/\pi\geq \Gamma((d+1)/2)/\sqrt{\pi}$ and that $\p/\p\beta_{\rl}\,r_\rl^2 \geq \p/\p\beta_\rl\,s_\rl^2$, we also find that \eqref{jineqc}, hence \eqref{jineq}, are satisfied.

\subsection{Ising chain in a transverse magnetic field}

The non-equilibrium steady state for energy transport in the Ising chain in a transverse magnetic field has been studied within the partitioning approach in \cite{ah,aschp,delucaising}. The Hamiltonian, in terms of Pauli matrices, takes the form
\beq
	H = -\frc12 \sum_n \lt(\sigma_n^x \sigma_{n+1}^x + h \sigma_n^z\rt)
\eeq
(we consider the case $h>1$ only). The right and left Hamiltonians $H_\rl$ and $H_\rr$ are defined by summing over $n\leq -1$ and $n\geq 1$, respectively, and $\delta H_{\rl\rr}$ is the term with $n=0$. The energy current density $\jj_n$ at the site $n=0$ is then evaluated by $\jj_0 = i[H,H_\rr - H_\rl]/2$.

 The model can be exactly solved using the Jordan-Wigner procedure, which maps the Ising spins to fermionic operators. In terms of local fermionic operators $\cc_n$ on sites $n\in\Z$, with $\{\cc_n,\cc_{n'}^\dag\} = \delta_{n,n'}$, the energy density and current are
\beqa
	\hh_n &=& -h \cc_n^\dag \cc_n + \frc12(\cc_n-\cc_n^\dag)(\cc_{n+1}+\cc_{n+1}^\dag) \n
	\jj_n &=& \frc{ih}2 \lt(\cc_n^\dag \cc_{n+1} - \cc_{n+1}^\dag \cc_n\rt).
\eeqa
One can explicitly verify that the energy current is the density of a conservation law:
\beq
	i[H,\jj_n] + \uu_{n+1}-\uu_n = 0
\eeq
with
\beq
	\uu_n = \frc h2 \lt(-\cc_n^\dag \cc_n + \frc12(\cc_{n-1}^\dag -\cc_{n-1})(
	\cc_{n+1}^\dag + \cc_{n+1})\rt).
\eeq
The diagonalization is performed by the transformation
\beq
	\cc_n = (-1)^n \int_{-\pi}^\pi \frc{\dd\theta}{2\pi}\,e^{in\theta}
	\lt(\cos\phi(\theta)\, A(\theta) - i \sin\phi(\theta) \,A^\dag(-\theta)\rt)
\eeq
where $\phi(\theta)\in[-\pi,\pi]$ is defined by
\beq
	\ep(\theta) e^{2i\phi(\theta)} = e^{i\theta}-h,\quad \ep(\theta)>0
\eeq
and $\{A(\theta),A^\dag(\theta')\} =2\pi \delta(\theta-\theta')$. The Hamiltonian is then expressed as
\beq
	H = \sum_n \hh_n = \int_{-\pi}^\pi \frc{\dd\theta}{2\pi}\,\ep(\theta) A^\dag(\theta) A(\theta)
\eeq
where $\ep(\theta) = \sqrt{h^2+1-2h\cos\theta}$.

It was shown in \cite{ah,aschp,delucaising} that, as in the Klein-Gordon case, the steady-state corresponds to independent thermalization of right and left movers; the density matrix is
\beq
	\rho_\rs = e^{-\int_{-\pi}^\pi \frc{\dd\theta}{2\pi}\,W(\theta)A^\dag(\theta) A(\theta)}
\eeq
where $W(\theta) = (\beta_\rl \Theta(\theta) + \beta_\rr \Theta(-\theta))\,\ep(\theta)$. One can then evaluate averages using the formula
\beq
	\sta{A^\dag(\theta) A(\theta')} =  \frc{2\pi\delta(\theta-\theta')}{
	1+e^{W(\theta) \ep(\theta)}}.
\eeq
One finds
\beqa\label{isingsta}
	\sta{\jj} &=& \int_{-\pi}^\pi \frc{\dd\theta}{2\pi} \,\frc{h\sin\theta}{1+e^{W(\theta)}} \n
	\gen{\uu}_{\rl,\rr} = \gen{\uu}_{\beta_{\rl,\rr}} &=& \int_{-\pi}^\pi \frc{\dd\theta}{2\pi}  \frc{h^2 \sin^2\theta}{\ep(\theta)\lt(1+e^{\beta_{\rl,\rr}\ep(\theta)}\rt)}.
\eeqa

In order to verify \eqref{jineq} and  \eqref{jineqc}, we need to evaluate the Lieb-Robinson velocity $v$. This can be calculated as the maximal group velocity $\dd\ep(\theta)/\dd \theta = h\sin\theta/\ep(\theta)$ over all values of wave numbers $\theta$ (we will come back to why this is the Lieb-Robinson velocity in a future work):
\beq
	v = {\rm max} \lt(\frc{h\sin\theta}{\ep(\theta)}:\theta\in[-\pi,\pi]\rt).
\eeq
The maximum at $\theta=\theta^*$ is obtained by solving
$\dd \lt(h\sin\theta/\ep(\theta)\rt)/\dd\theta|_{\theta=\theta^*}=0$, and we find
$\cos\theta^* = 1/h$
and
 $\sin\theta^* = \sqrt{1-1/h^2}$.
The result 
is
$v = 1$, hence \eqref{jineq} and \eqref{jineqc} do not involve additional velocity factors.

We may then verify \eqref{jineqc}, which implies \eqref{jineq}, by evaluating the derivatives $-\p\sta{j}/\p\beta_\rl$ and $-\p\gen{\uu}_{\beta_\rl}/\p\beta_\rl$. Using \eqref{isingsta}, it is sufficient to check that
\beq
	h\sin\theta \geq \frc{h^2\sin^2\theta}{\ep(\theta)}\quad\forall\;\theta\in[0,\pi].
\eeq
One can indeed show that $\ep(\theta)>h\sin\theta$: the inequality is immediate at $\theta=0,\pi$, both sides are continuous, and both side are equal to each other if and only if $\theta = {\rm arccos}\,(1/h)$.

\subsection{Massive integrable quantum field theory}

Integrable models of QFT with single particle spectra have also been studied within the partitioning approach, with conjectured exact expressions for the current using a non-equilibrium generalization \cite{doyonint} of the thermodynamic Bethe ansatz \cite{tba1}. The main object is a  (dimensionless) ``free energy''
\beq\label{fa}
	f_{a}(\beta_\rl,\beta_\rr) = -\int \frc{\dd\theta}{2\pi}\,m\cosh\theta\,\log\lt(
	1+e^{-\varep_{a}(\theta)}\rt).
\eeq
The function $\varep_a(\theta)$ solves the integral equation
\beqa\label{ea}
	\varep_a(\theta) &=& W(\theta) + am\sinh(\theta) - \\
	&& \int \frc{\dd\gamma}{2\pi}
	\,\varphi(\theta-\gamma) \log\lt(1+e^{-\varep_a(\gamma)}\rt) \no
\eeqa
where $W(\theta)$, as above, is the driving term representing the non-equilibrium steady state, and $\varphi(\theta)$ is the scattering phase of the two-particle process. The parameter $a$ generates the connected correlation functions of the energy current (momentum density) operator $ \jj=\TT^{01}$; for instance, $\lt.\frc{\p}{\p a} f_a(\beta_\rl,\beta_\rr)\rt|_{a=0} =\sta{\jj}$.
The equilibrium free energy is $f(\beta)=\beta^{-1} f_0|_{\beta_\rl=\beta_\rr=\beta}$, and the pressure is given by $\gen{\uu}_{\rl,\rr} = -f(\beta_{\rl,\rr})$. Hence for all relevant quantities in \eqref{jineq}, we obtain sets of integral equations which can be solved numerically. In \cite{doyonint} the current was analyzed, in particular, for the one-parameter family of ``roaming-trajectory'' models \cite{zamoroam}, with scattering phase given by $\varphi(\theta) = {\rm sech}\,(\theta-\theta_0) + {\rm sech}\,(\theta+\theta_0)$ where $\theta_0\in\R^+$ (the sinh-Gordon model at the self-dual point is obtained by setting $\theta_0=0$). We have performed a numerical analysis, and found that, for values of $\beta_\rl$ and $\beta_\rr$ far enough from each other, inequality \eqref{jineq} is broken. We hope to investigate this in future works.

\section{Proof of a relativistic thermodynamic relation} \label{proofrel}

We prove (\ref{addr}) as follows, focussing on $d=1$ for simplicity. Let us denote $\TT^{00}=\hh$ and $\TT^{01}=\pp$, and $\genc{{\tt a}{\tt b}} = \gen{{\tt a}{\tt b}} - \gen{{\tt a}}\gen{{\tt b}}$. We assume that connected correlation functions decay fast enough. We need to show
\beq\label{add}
	\int \dd x\,\genc{\hh(x)\uu(0)} = \int \dd x\,\genc{\pp(x)\jj(0)}.
\eeq
By Poincar\'e invariance, the stress-energy tensor is conserved and symmetric, hence $\p_t \hh = - \p_x \pp$. Let $B_\ell := i\int_{-\ell}^\ell \dd x\,x\hh(x)$. Recalling that $[P,\Or] = i\p_x \Or$ and $[H,\Or] = -i\p_t\Or$, we have
\beqa
	[P,B_\ell] &=& -\int_{-\ell}^\ell \dd x\,x\p_x\hh(x) = H_\ell - \ell\,\hh_s(\ell)\n{}
	[H,B_\ell] &= &-\int_{-\ell}^\ell \dd x\,x\p_x\pp(x) = P_\ell - \ell\,\pp_s(\ell)
	\no
\eeqa
where $\hh_s(\ell):=\hh(\ell)+\hh(-\ell)$, $\pp_s(\ell):=\pp(\ell)+\pp(-\ell)$,  $H_\ell: = \int_{-\ell}^\ell \dd x\,\hh(x)$ and $P_\ell: = \int_{-\ell}^\ell \dd x\,\pp(x)$. Hence,
\beqa
	-\gen{B_\ell[P,\uu(0)]} &=& \gen{[P,B_\ell]\uu(0)} = \gen{(H_\ell-\ell\,\hh_s(\ell))\uu(0)}\n &=& \genc{H_\ell\uu(0)} + \ell\genc{\hh_s(\ell)\uu(0)}
\eeqa
where we used $\gen{H_\ell} = \ell\gen{\hh_s(\ell)}$ by translation invariance. With decay faster than $1/\ell$, taking the large-$\ell$ limit we obtain $-\gen{B_\ell[P,\uu(0)]} = \int \dd x\,\genc{\hh(x)\uu(0)}$. A similar calculation shows that $-\gen{B_\ell[H,\jj(0)]} = \int \dd x\,\genc{\pp(x)\jj(0)}$. Using \eqref{JI} we get \eqref{add}. Naturally, this holds for any $\uu$ and $\jj$ related by a conservation law. Hence with \eqref{qj} we also have
\beq
	-\frc{\p}{\p\beta} \gen{\jj}_{\beta,\nu,\t\mu^{(j)}} =
	\frc{\p}{\p\nu} \gen{\qq}_{\beta,\nu,\t\mu^{(j)}}.
\eeq

\end{document}